\documentclass[aps,prd,twocolumn,amsmath,nofootinbib]{revtex4-1}
\usepackage{amssymb,amsmath}
\usepackage{graphicx}
\usepackage{color}
\usepackage{xfrac}

\usepackage[caption=false]{subfig}

\usepackage{ulem}
\usepackage[dvipsnames]{xcolor}

\begin{document}
\title{Comment on ``Analog Schwarzschild-like geometry in fluids with
external pressure''}

\author{Jo\~ao Paulo M. Pitelli}
\email[]{pitelli@unicamp.br}
\affiliation{Departamento de Matem\'atica Aplicada, Universidade Estadual de Campinas, 13083-859, Campinas, S\~ao Paulo, Brazil}

\author{Ricardo A. Mosna}
\email[]{mosna@unicamp.br}
\affiliation{Departamento de Matem\'atica Aplicada, Universidade Estadual de Campinas, 13083-859, Campinas, S\~ao Paulo, Brazil}

\author{Christyan C. de Oliveira}
\email[]{chris@ifi.unicamp.br}
\affiliation{
Instituto de F\'isica ``Gleb Wataghin'' (IFGW), Universidade Estadual de Campinas, 13083-859 Campinas, SP, Brazil}

\author{Mauricio Richartz}
\email[]{mauricio.richartz@ufabc.edu.br}
\affiliation{
Centro de Matemática, Computação e Cognição, Universidade Federal do ABC (UFABC), 09210-170 Santo Andr\'e, SP, Brazil}

\begin{abstract}

In Ref.~\cite{de oliveira}, exact (not only conformally related) analogue models for the Schwarzschild and Reissner-Nordstr\"om spacetimes were found. The background non-relativistic fluid flow was sustained by an external body force which is not affected by the linearised fluctuations. Following a different route, by modelling the external force as the gradient of an external pressure, it was shown in~\cite{bilic} that the speed of sound is significantly modified. This effective speed of sound turns out to be inconsistent with the continuity equation. In this comment we analyse these two contradictory conclusions. We first show that the heuristic justification for the introduction of the external force via a variational principle given in Ref.~\cite{bilic} is conceptually incorrect. Then, by adding the external force appropriately,  we show that the conclusions in~\cite{de oliveira} remain valid.
\end{abstract}

\maketitle
\section{Perfect Fluids from a variational principle}
The usual relativistic description of perfect fluids consists in defining the energy momentum tensor
\begin{equation}
T_{\mu\nu}^{\textrm{fluid}}=(\rho+p)U_{\mu}U_{\nu}+p \eta_{\mu\nu},
\label{energy-momentum}
\end{equation}
along with the conservation equation
\begin{equation}
T^{\mu\nu}_{\textrm{fluid},\nu}=0.
\label{conserved}
\end{equation}
In Eq.~(\ref{energy-momentum}),  $\rho$ is the density of total mass-energy, $p$ is the fluid pressure, $\eta_{\mu\nu}$ is the Minkowksi metric and $U^{\mu}(x)$ is the four-velocity field of the fluid.  Eq.~(\ref{conserved}) leads to  the dynamical equations of motion of the fluid.

In a seminal paper~\cite{schutz},  Schutz found an equivalent set of equations for the fluid via a  variational principle through the action
\begin{equation}
S_{\textrm{fluid}}=\int{pd^4x}.
\label{action}
\end{equation} 
In Schutz' formalism,  the four-velocity is decomposed in terms of six potentials
\begin{equation}
U_{\nu}=\mu^{-1}(\phi_{,\nu}+\alpha\beta_{,\nu}+\theta s_{,\nu}), 
\label{four-velocity}
\end{equation}
where the field $\mu$ is the specific enthalpy while $s$ is the specific entropy.  Schutz also  assumed  an equation of state of the form $p=p(\mu,s)$ together with the first law of thermodynamics in the form
\begin{equation}
dp=\rho_0 d\mu-\rho_0 Tds,
\label{first law}
\end{equation}
where $\rho_0$ is the rest mass density of the fluid and $T$ is the temperature of the fluid.  The enthalpy $\mu=(\rho+p)/\rho_0$ is not an  independent field, since the normalization equation $U^{\nu}U_{\nu}=-1$ implies
\begin{equation}
\mu^2=-\eta^{\nu\lambda}(\phi_{,\nu}+\alpha\beta_{,\nu}+\theta s_{,\nu})(\phi_{,\lambda}+\alpha\beta_{,\lambda}+\theta s_{,\lambda}).
\label{normalization}
\end{equation}
The remaining five potentials in Eq.~(\ref{four-velocity})  have their own evolution equations.

The variation of the action given by Eq.~(\ref{action}) with respect to  $\phi$, $\theta$, $s$, $\alpha$ and $\beta$ gives the following evolution equations (Eqs.~(\ref{first law}) and~(\ref{normalization}) are extensively used in this derivation - see Ref.~\cite{schutz} for details)
\begin{equation}\left.\begin{array}{ccc}
\delta S_{\textrm{fluid}}/\delta \phi = 0 &\Rightarrow & (\rho_0 U^{\nu})_{,\nu}=0, \\
\delta S_{\textrm{fluid}}/\delta s = 0 &\Rightarrow & \ \ \ \  U^{\nu}\theta_{,\nu}=T,\\
\delta S_{\textrm{fluid}}/\delta \beta = 0 &\Rightarrow & U^{\nu}\alpha_{,\nu}=0,\\
\delta S_{\textrm{fluid}}/\delta \theta = 0 &\Rightarrow & \ \ \ \  U^{\nu}s_{,\nu}=0,\\
\delta S_{\textrm{fluid}}/\delta \alpha = 0 &\Rightarrow & U^{\nu}\beta_{,\nu}=0.
\label{equations}
\end{array}\right.
\end{equation}
Notice that the first and the fourth equations in~(\ref{equations}) imply that mass is conserved and that the fluid is isentropic, respectively. Finally, the normalization condition and the last two equations in~(\ref{equations}) imply the evolution equation for the remaining potential $\phi$,
\begin{equation}
U^{\nu}\phi_{,\nu}=-\mu.
\label{eq phi}
\end{equation}

Clearly, Eqs.~(\ref{equations})  and~(\ref{eq phi}) only make sense if they correspond to the correct equations of motion for the fluid, i.e., if they are equivalent to Eq.~(\ref{conserved}). In fact, Schutz showed  that this is true.  In this way,   the action~(\ref{action}), along with the velocity decomposition given by Eq.~(\ref{four-velocity}) and the first law of thermodynamics~(\ref{first law}) give the correct prescription for a perfect fluid.

\section{external force from an external potential?}
A perfect fluid subjected to  an external applied force satisfies the following equation, 
\begin{equation} \label{forcex}
T^{\mu\nu}_{\phantom{\mu\nu};\nu}=\mathcal{F}^{\mu},
\end{equation}
where $\mathcal{F}^{\mu}$ is the associated four-force density. Two important questions regarding the source $\mathcal{F}^{\mu}$ arise, namely, what is the origin of $\mathcal{F}^{\mu}$ and how do we control this external force?

 In Ref.~\cite{bilic}, $\mathcal{F}^{\mu}$ is considered as the four-gradient of an external pressure $p^{\text{ext}}$, i.e., 
 \begin{equation}
T^{\mu\nu}_{\phantom{\mu\nu};\nu}=p^{\textrm{ext},\mu}.
\label{external pressure}
\end{equation}
The function $p^{\textrm{ext}}$ could, in principle, be arbitrarily given by an external agent. This is motivated by the least action principle for a  perfect fluid through the addition of an extra  potential $U(x)$ to the fluid Lagrangian density.  This potential is assumed  to be independent of any  dynamical field.  The action in this case becomes
\begin{equation}
S_{\textrm{modified}}=\int{\left[p(\phi,\alpha,\beta,\theta,s,g^{\mu\nu})-U(x)\right]\sqrt{-g}d^4x},
\label{modified}
\end{equation}
where we emphasize the $p$ dependence on the independent dynamical fields. The  external potential plays the role of (minus) an external pressure $p^{\textrm{ext}}=-U$.  Then, it is argued that this external pressure plays the role of an external force on the background fluid.

In fact,  the potential $U(x)$ modifies the stress-energy tensor.  A straightforward calculation (see Ref.~\cite{schutz} for the details) shows that the variation of the action in Eq.~(\ref{modified}) gives the modified stress-energy tensor  
\begin{equation}
T_{\mu\nu}=T_{\mu\nu}^{\textrm{fluid}}+p^{\textrm{ext}}g_{\mu\nu},
\label{new stress}
\end{equation}
where $T_{\mu\nu}^{\textrm{fluid}}$ is given by Eq.~(\ref{energy-momentum}). However,  since $p^{\textrm{ext}}=-U(x)$ does not depend on the dynamical variables, it does not modify the evolution equations in Eq.~(\ref{equations}).  These equations (as shown in Ref.~\cite{schutz}) are equivalent to the usual equation of motion $T^{\nu\lambda}_{\textrm{\tiny{fluid}};\lambda}=0$.  Hence, the stress-energy tensor in Eq.~(\ref{new stress}) is not conserved, unless  $p^{\textrm{ext}}$ is constant. This leads to the energy-momentum tensor of two noninteracting fluids, one of them with equation of state $p=-\rho$, which represents a cosmological constant.

In order to have an effective interaction, in Eq.~(\ref{modified}) we should add a potential of the form $U=U(\phi,\alpha,\beta,\theta,s,g^{\mu\nu},t)$.  But this would dramatically change the Euler-Lagrange equations. Hence, we conclude that Eq.~(4) in Ref.~\cite{bilic}, which states that
\begin{equation}
T^{\mu\nu}_{\phantom{\mu\nu};\nu}=0\Leftrightarrow T^{\mu\nu}_{\textrm{fluid};\nu}=p^{\textrm{ext},\mu}
\label{incorrect}
\end{equation}
cannot be obtained from a variational principle unless  $p^{\textrm{ext},\mu}=0$. Consequently, both questions in the beginning of this section remain unanswered.

\section{External force from Newton's second  law}
But how can we model an external  force acting on a perfect fluid in a consistent way? To answer this question, we go back to the well-known subject of classical mechanics of particles and look at the generalization of Newton's second law for a relativistic particle~\cite{mihalas}.  In inertial coordinates with metric
\begin{equation}
ds^2=-dt^2+d\vec{x}^2,
\end{equation}
we have
\begin{equation}
F^{\alpha}=\frac{d}{d\tau}(m_0 V^{\alpha}),
\label{newton}
\end{equation} 
with $V^{\alpha}=\gamma(1,\vec{v})$ being the four-velocity of the particle, $\gamma=1/\sqrt{1-\vec{v}^2}$ and $m_0$ the particle's rest mass. The left hand side of Eq.~(\ref{newton}) is the external force we actually control, while the right hand side is the particle response to this force.  As in \cite{visser1998}, from now on, we assume that there is no back-reaction, i.e.~the individual particles  can be considered as test particles so that their dynamics do not affect the strong potential responsible for $F^{\alpha}$. Let us also suppose that we are dealing with an  ordinary body force which does not change the particle's internal state (this is equivalent to $m_0=\textrm{constant}$). This implies (with the help of the normalization condition $V^{\alpha}V_{\alpha}=-1$)
\begin{equation}
V^{\alpha}F_{\alpha}=0
\label{good force}
\end{equation}
so that
\begin{equation}
F^{\alpha}=\gamma(\vec{F}\cdot \vec{v},\vec{F}),
\label{four force}
\end{equation}
with $\vec{F}=\frac{d\vec{p}}{dt}$ and $\vec{p}=m_0\gamma \vec{v}$.

In this way, the four-force density on a proper volume $\delta V_0$ of the fluid is given by
\begin{equation}
\mathcal{F}^{\alpha}=NF^{\alpha}/\delta V_0,
\end{equation}
where $N$ is the number of fluid elements in the volume $\delta V_0$.

If we define the three-force density as
\begin{equation}
\vec{\mathcal{F}}= N\vec{F}/\delta V,
\label{force density}
\end{equation}
where $\delta V$ is a volume element on an arbitrary frame, we have 
\begin{equation} 
\mathcal{F}^{\alpha}=\gamma(\delta V/\delta V_0)(\vec{\mathcal{F}}\cdot \vec{v},\vec{\mathcal{F}})=(\vec{\mathcal{F}}\cdot \vec{v},\vec{\mathcal{F}}),
\end{equation}
where we used the relativistic relation $\delta V_0=\gamma \delta V$.
Notice that, although we can control the vector%
\footnote{As an example, consider an electric charged fluid. In this case, $\vec{F}$ would be equal to $q\vec{E}$, where $q$ is the electric charge of each fluid component and $\vec{E}$ is an external electric field.}
 $\vec{F}$ in Eq.~(\ref{force density}), the force density $\vec{\mathcal{F}}$ will be coupled to the fluid density $\rho$ through  Eq.~(\ref{force density}). This difference between  the external applied force and the fluid response to this force will be crucial in what follows.

We generalize the relativistic version of Newton's second law given by Eq.~(\ref{newton})  through the following relation%
\footnote{A comparison between Eqs.~(\ref{generalization}) and (\ref{external pressure}) shows our main point in this comment, namely, while the right hand side of Eq.~(\ref{generalization}) depends  on the fluid density, the right hand side of Eq.~(\ref{external pressure}) is supposed to be external and independent of the fluid configuration.}
\begin{equation}
T^{\mu\nu}_{\textrm{fluid},\nu}=\mathcal{F}^\mu.
\label{generalization}
\end{equation}

By projecting Eq.~(\ref{generalization}) with the four-velocity vector $U^{\alpha}$ and using Eq.~(\ref{good force}) we have
\begin{equation}
\rho_{,\nu}U^{\nu}+(\rho+p)U^{\nu}_{\phantom{\nu},\nu}=0.
\end{equation}
In the non-relativistic limit ($v\ll1$ and $p\ll\rho$), this leads to the conservation equation
\begin{equation}
\frac{\partial\rho_0}{\partial t}+\nabla\cdot(\rho_0\vec{v})=0. 
\end{equation}

Finally, the non-relativistic limit of Eq.~(\ref{generalization}) contracted with the projection tensor
\begin{equation}
P^{\sigma}_{\phantom{\sigma}\nu}=\delta^{\sigma}_{\phantom{\sigma}\nu}+U^{\sigma}U_{\nu}
\end{equation} 
leads to Euler's Equation
\begin{equation}
\rho_{0}\left[\frac{\partial\vec{v}}{\partial t}+\vec{v}\cdot\nabla\vec{v}\right]=\nabla p+\vec{\mathcal{F}}.
\end{equation}
In the non-relativistic limit, $\vec{\mathcal{F}}=N\vec{F}/\delta V_0$ and $\rho_0=m_0 N/\delta V_0$. If, in addition, the force $\vec{F}$ could be derived from an external potential such that $\vec{F}\equiv m_0\nabla \phi$, then
\begin{equation}
\frac{\partial\vec{v}}{\partial t}+\vec{v}\cdot\nabla\vec{v}=\frac{\nabla p}{\rho_0}+\nabla\phi.
\label{correct}
\end{equation}
This is the usual Euler's equation subjected to an external potential $\phi$. This potential   is, clearly, not affected by any perturbation on the fluid background.

Let us compare Euler's equation given by Eq.~(\ref{correct}) with the one found in  Ref.~\cite{bilic}. By assuming the validity of Eq.~(\ref{incorrect}),  we have (in the non-relativistic limit)
\begin{equation}
\frac{\partial\vec{v}}{\partial t}+\vec{v}\cdot\nabla\vec{v}=\frac{\nabla p}{\rho_0}+\frac{\nabla p^{\textrm{ext}}}{\rho_0}.
\label{incorrect2}
\end{equation}
In this case, the last term on the right hand side of Eq.~(\ref{incorrect2})  would be  affected by a linear perturbation in the fluid background (specifically by the density perturbation $\rho_0\to\rho_0+\delta\rho_0$). In particular,  the non-relativistic specific enthalpy given by $h$, with $\nabla h=\nabla p/\rho_0$, should be replaced by an effective enthalpy given by $h+V$, with $\nabla V=\nabla p^{\textrm{ext}}/\rho_0$. This would change the speed of sound significantly. It is precisely this modification that leads Ref.~\cite{bilic} to a different result than Ref.~\cite{de oliveira}.

\section{Conclusions}

The  external potential in the action, Eq.~(\ref{modified}), does not correspond to an interaction between the fluid and an external agent. The correct way to introduce such an interaction would require a field dependent potential of the form $U=U(x,\phi, \alpha,\beta,\theta,s,t)$. However, this extra potential would certainly affect the equations of motion given by Eq.~(\ref{equations}), possibly invalidating the equivalence between these equations and the perfect fluid description.  In fact,  the authors are unaware of any consistent way of adding an external force via a variation principle in the perfect fluid context.  

In Ref.~\cite{de oliveira}, we introduced an external force directly on the right hand side of Euler's equation.  This is equivalent to applying the relativistic version of Newton's second law to a continuum of particles that define the fluid. By assuming a strong external potential,  we conclude that it is possible to model a perfect fluid subject to an external body force which is not affected by the fluid dynamics, preserving the usual speed of sound $c_s^2=\frac{\partial p}{\partial \rho}$ as in Ref.~\cite{de oliveira}.

\acknowledgements
The authors are indebted to A. Saa for clarifying several questions during the development of this comment. J.P.M.P. was partially supported by Conselho Nacional de
Desenvolvimento Científico e Tecnológico (CNPq, Brazil) under Grant No.
311443/2021-4. R. A. M. was partially supported by Conselho Nacional de Desenvolvimento Científico e Tecnológico (CNPq, Brazil) under Grant No. 310403/2019-7. C. C. O. acknowledges support from the Conselho Nacional de Desenvolvimento Cient\'{i}fico e Tecnol\'{o}gico (CNPq, Brazil), Grant No. 142529/2018-4. M. R. acknowledges support from the Conselho Nacional de Desenvolvimento Científico e Tecnológico (CNPq, Brazil),
Grant No. 315664/2020-7.

\end{document}